\documentclass[12pt, a4paper]{iopart}
\usepackage{iopams}

\usepackage{graphicx}

\begin{document}
\title{Spin polarized tunneling in MgO-based tunnel junctions with superconducting electrodes}
\author{Oliver Schebaum}
\address{Bielefeld University, Thin films and Physics of Nanostructures, 33615 Bielefeld}
\author{Jagadeesh S. Moodera}
\address{Francis Bitter Magnet Lab., Mass. Inst. of Tech., 02139 Cambridge, USA}
\author{Andy Thomas}
\ead{andy.thomas@uni-bielefeld.de}
\address{Bielefeld University, Thin films and Physics of Nanostructures, 33615 Bielefeld}
\begin{abstract}
We prepared magnetic tunnel junctions with one ferromagnetic and one superconducting Al-Si electrode. Pure cobalt electrodes were compared with a Co-Fe-B alloy and the Heusler compound Co$_2$FeAl. The polarization of the tunneling electrons was determined using the Maki-Fulde-model and is discussed along with the spin-orbit scattering and the total pair-breaking parameters. The junctions were post-annealed at different temperatures to investigate the symmetry filtering mechanism responsible for the giant tunneling magnetoresistance ratios in Co-Fe-B/ MgO/ Co-Fe-B junctions.
\end{abstract}
%
\tableofcontents

\section{Introduction}
The discovery of the giant magnetoresistance effect in 1988 \cite{Baibich:1988p14033, Binasch:1989p14034}, which led to the realization of room temperature tunnel magnetoresistance (\textsc{tmr}) of more than a few precent \cite{Moodera:1995p66, Miyazaki:1995p5737} and the 2007 Nobel prize in Physics for Gr\"unberg and Fert, has led to intensive research in the field of spinelectronics or spintronics \cite{Prinz:1998p5}. In the past years, this research has been driven by larger and larger \textsc{tmr}-ratios in alumina magnetic tunnel junctions (\textsc{mtj}s), which have now reached up to 80\% at room temperature \cite{Wei:2007p60}.

Recently, the search for new material combinations has resulted in increased interest in MgO-based epitaxial magnetic tunnel junctions, where large \textsc{tmr}-ratios were predicted and observed due to the so-called ``symmetry filtering'' \cite{Butler:2001p4537, Mathon:2001p6616, Parkin:2004p71, Yuasa:2004p106}. Junctions with \textsc{tmr}-ratios of up to 600\% have been realized in Co-Fe-B/ MgO systems \cite{Ikeda:2008p3496}. Along this line another system of materials of interest is the class of Heusler compounds, which are predicted to be half-metallic in some cases \cite{DeGroot:1983p3287}.

Spin polarized tunneling (\textsc{spt}) from an electrode into superconducting aluminum was pioneered by Meservey and Tedrow in 1970 \cite{Meservey:1970p77} and can be seen as a complementary technique to \textsc{tmr} experiments. \textsc{spt} measurements would be a more direct way to measure the spin polarization ($P$) of tunneling electrons coming from the ferromagnet than the \textsc{tmr} experiments. If we look only at the \textsc{tmr} ratio and pay no attention to the \textsc{spt} measurements, the one half of the world cannot understand the pleasures of the other. However, only a few experiments with MgO based systems have been reported \cite{Parkin:2004p71, Kant:2004p7468, Yang:2006p34, Yang:2007p35}. In this work, we present \textsc{spt}-experimental results on MgO-based junctions with ferromagnetic electrodes such as Cobalt, Co-Fe-B and Co$_2$FeAl (Heusler).
\section{Preparation}
The samples investigated in the present study were prepared using DC- and RF-magnetron sputtering in an automatic sputtering system at ambient temperature. The tunnel junctions were made using molybdenum shadow masks with a cross-strip geometry. The 300\,$\mu$m width of the strips results in a junction area of approximately $9\times10^{4}$\,$\mu$m$^2$. To avoid short circuits at the edges of the stripes, the tunnel barrier was sputtered without a shadow mask. After deposition the samples were annealed ex-situ in a vacuum furnace at various temperatures for one hour. The base pressure of both the sputtering system and the vacuum furnace was $1\times 10^{-7}$\,mbar.

The differential conductance, $\mathrm{d}I/\mathrm{d}V$ versus bias voltage $V$ of the S-I-F-tunnel junctions was measured with standard lock-in techniques. The AC excitation voltage was always kept below the thermal smearing threshold $k_B \cdot T$. The measurements were performed in $^3$He cryostats. The Al-Si--MgO--Co samples were measured at the Francis Bitter Magnet Laboratory at a temperature of $0.45$\,K. All other samples were investigated at Bielefeld University at $0.3$\,K. While performing the measurements, a magnetic field was applied in the plane of the sample to create Zeeman splitting of the superconducting density of states \cite{Meservey:1970p77}.

The transition temperature of the Al-Si--MgO--Co tunnel junction was measured by passing a constant current of $10\,\mu A$ through the Al-Si strip while measuring the voltage drop across this strip. For the other junctions, the transition temperature of the tunnel junction was determined by measuring the differential conductance $\mathrm{d}I/\mathrm{d}V$ at $V=0$ while cooling the sample. At the transition to the superconducting phase, a decrease in the conductance was visible due to the formation of the superconducting energy gap. The error in $T_c$ is $\pm 0.02$\,K for both methods.

\section{Fitting procedure}
The spin polarization, $P$, of the tunneling current was determined by fitting theoretical curves calculated using the Maki-Fulde theory for a superconductor in a magnetic field \cite{Maki:1964p6983, Fulde:1973p9258} to the measured conductances \cite{Alexander:1985p8294}. In addition to $P$, the parameters for this theory include the measuring temperature $T$, the transition temperature $T_c$, the magnetic field $H$, the spin-orbit scattering parameter $b$ and the pair breaking parameter $\zeta(H)$ \cite{Meservey:1994p4, Worledge:2000p2534}. For the calculation of the theoretical Maki curves, $H$, $T$ and $T_c$ were kept at their known or measured values and not varied as fitting parameters. The parameters $b$ and $P$ were chosen to fit the curves for the three different magnetic fields  with the same set of parameters. The Maki parameters are shown at the corresponding curves in the figures.

\section{Cobalt-based junctions}
First, we present the measurements from cobalt junctions with an Al-Si (3.84\,nm)/MgO (1.8\,nm)/Co (20\,nm) layer stack. The results of two different systems are presented: i) as-prepared and ii) ex-situ annealed at a temperature of $T_a =  325\,^{\circ}$C. The critical temperatures of Al-Si were $2.21 \pm 0.12$\,K and $2.41 \pm 0.05$\,K, respectively. 

\begin{figure}\hfill
	\includegraphics{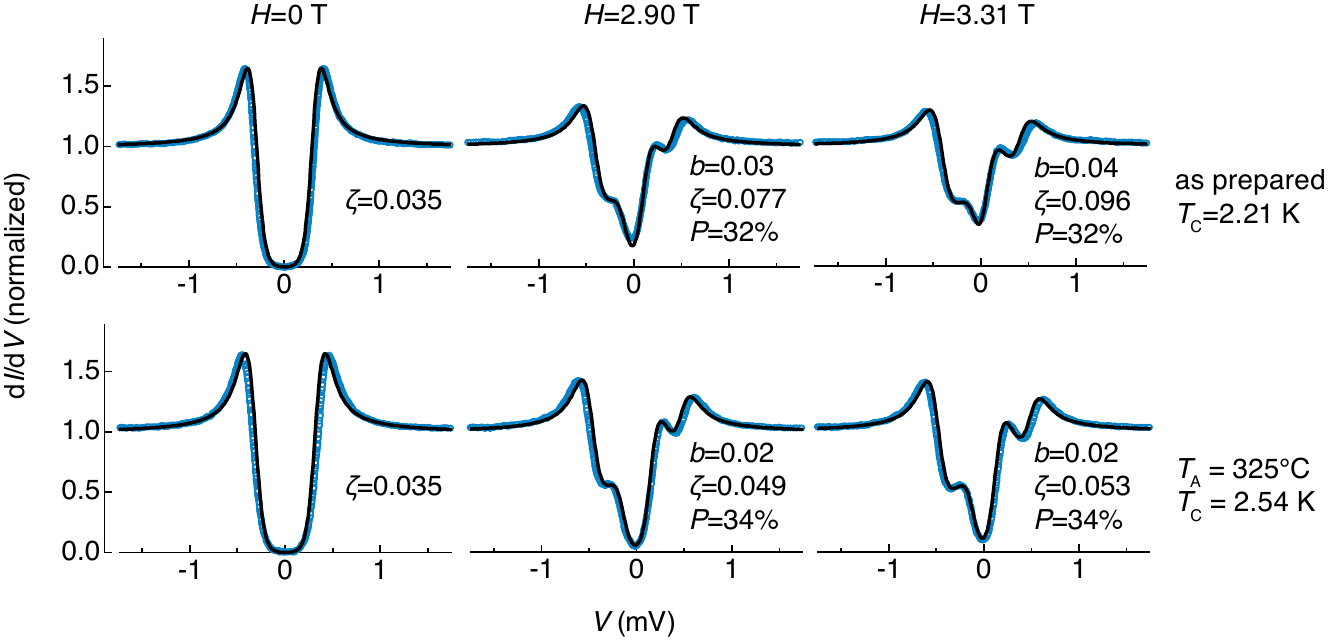}
	\caption{Differential conductivity of a Al$_{95}$Si$_5$ (3.84\,nm)/ MgO (1.8\,nm)/ Co (20\,nm) junction. The top row shows the as-prepared sample at different magnetic fields and the bottom row depicts the junction annealed at $325\,^\circ$C. The measurements and Maki calculations are displayed as blue circles and black lines, respectively.}
	\label{Co_SPT}
\end{figure}

The differential conductance measurements are depicted in Figure \ref{Co_SPT}. The Al-Si superconducting energy gaps and the Zeeman splits of Al-Si quasi particle density of states, as well as the asymmetry of the conductance curves, are apparent. The theoretical fit conductance curves for both samples show good agreement with the data. The spin-orbit scattering parameter, $b$, is between $0.02$ and $0.05$ for all measurements and is comparable to the published results for Al$_{98}$Si$_2$ alloys \cite{Parkin:2004p71}. The same reference reports pair-breaking parameters in a parallel magnetic field of $2$\,T between $\zeta = 0.024$ and $\zeta = 0.053$. If we extrapolate our parameters to match the magnetic field of $2$\,T, our layer stacks show values of $\zeta = 0.048$ and $\zeta = 0.042$ for the as-prepared and annealed samples, respectively. The lower $b$ values for the annealed sample are caused by the increased critical temperature.

The spin polarization of the as-prepared sample was determined to be $P = 32 \pm 2$\,\% and $34\pm2$\,\% for as grown and annealed junctions. This value matches the values of $P= 30 \pm 2$\,\% that have been reported by Kant et al.\ \cite{Kant:2004p7468}. Both values  appear to be low when compared to \textsc{tmr} values of epitaxial Co(001)/MgO(001)/Co(001) junctions. Those annealed junctions exhibit \textsc{tmr} ratios of 410\% at room temperature and 507\% at low temperatures \cite{Yuasa:2006p10780}. The high values can be explained by symmetry filtering \cite{Butler:2001p4537, Mathon:2001p6616}, which can only occur with ultra-thin, metastable, bcc Co(001) electrodes \cite{Bagayako:1983p10849}. However, the molecular beam epitaxy preparation of the junctions utilized by Yuasa et al.\ is incompatible with the preparation of \textsc{spt} samples \cite{Tanaka:1999p90, Parkin:2004p71, Kant:2004p10982, Worledge:2000p55}. 

\section{Co-Fe-B based junctions}
\textsc{Mtj}s with Co-Fe-B electrodes show the highest \textsc{tmr} ratios. This has been observed in alumina-based junctions where values of up to 80\% have been reached \cite{Wei:2007p60} and in MgO junctions where a ratio of more than 600\% at room temperature has been attained \cite{Ikeda:2008p3496}. The control \textsc{mtj}s that are produced in our group using the same sputtering systems as those in the present study reach TMR ratios of 73\,\% and 330\,\% for alumina and MgO tunnel barriers, respectively. The sputtered Co-Fe-B layers are initially amorphous, which is a result of the approximately 20\% boron content. A post-annealing temperature of typically more than $250\,^\circ$C leads to a crystallization of the layers \cite{Paluskar:2006p26}. 

At least two possible mechanisms have been proposed that give rise to the high \textsc{tmr}-ratios of the Co-Fe-B alloys. First, the smooth surface of the Co-Fe-B electrodes may improve the layer growth \cite{Yuasa:2007p18}. Second, the boron presence may increase the polarization of the tunneling electrons \cite{Paluskar:2008p23}. These were also the motivation for our \textsc{spt} investigation of Co-Fe-B, complimenting with our \textsc{tmr}-studies.

\begin{figure}\hfill
	\includegraphics{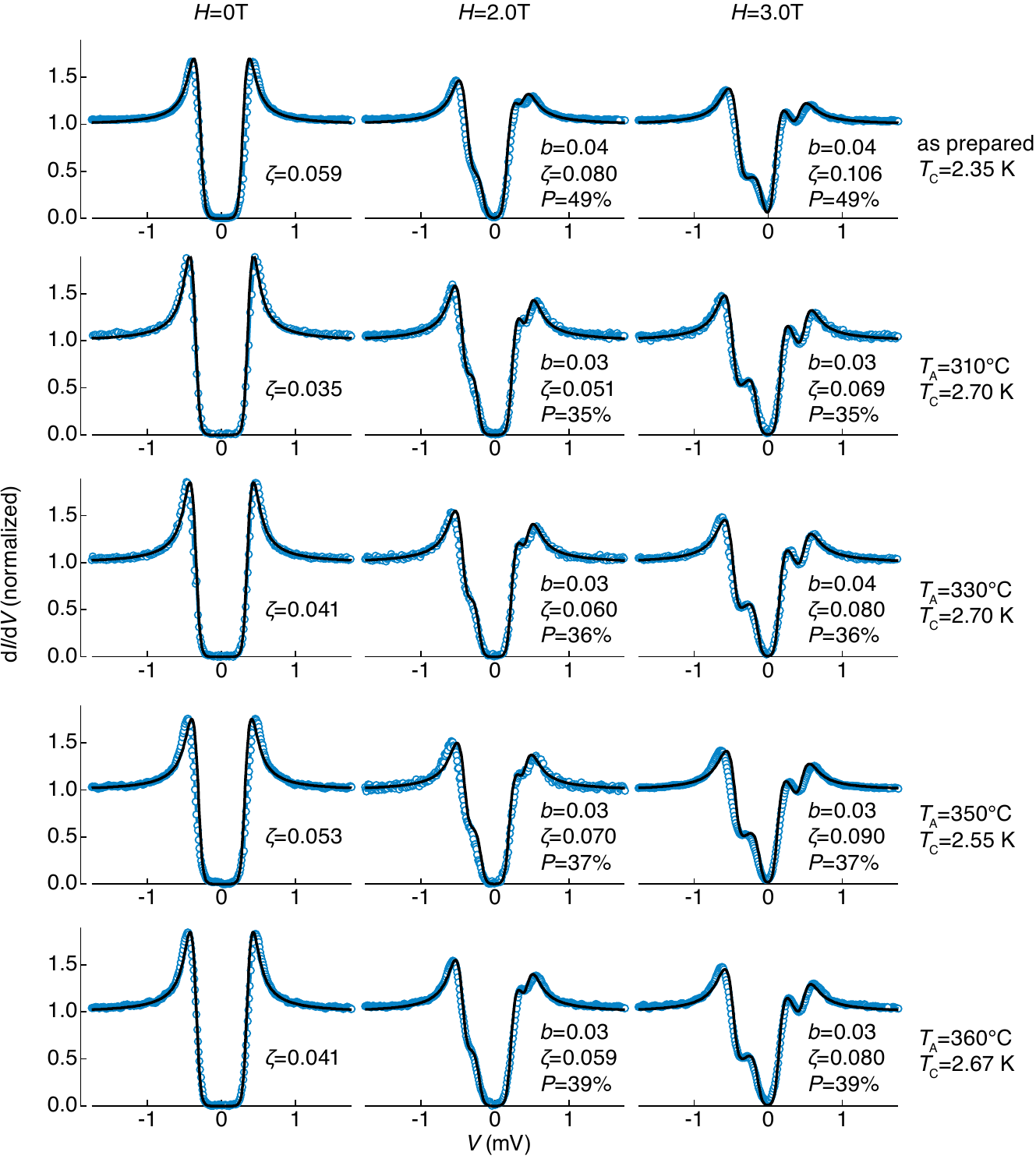}
	\caption{Differential conductance measurements of Al$_{95}$Si$_5$ (4.0\,nm)/ MgO (2.1\,nm)/ Co$_{40}$Fe$_{40}$B$_{20}$(20\,nm) junctions at different external magnetic fields and after various post-annealing temperatures.}
	\label{CoFeB_SPT}
\end{figure}

Figure \ref{CoFeB_SPT} displays the differential conductance measurements of the Co-Fe-B samples. Once again the \textsc{bcs} gap in the Al-Si density of states and its asymmetric spin split peaks in a magnetic field are visible. The calculated Maki curves are a good fit for the experimental data. The Al-Si $T_{\mathrm{c}}$ for the annealed samples is higher than for the as-prepared samples. Again, the spin-orbit scattering parameter and the pair breaking parameter are consistent with the values previously reported in the literature \cite{Parkin:2004p71}. However, it is interesting to observe the change in the magnitude of the spin polarization values with annealing temperature. The as-prepared sample starts with a ratio of 49$\pm$2\%, initially drops to 35$\pm$2\% at an annealing temperature of $T_a=310\,^\circ$C and continues with a slow increase to 39$\pm$2\% for $T_a=360\,^\circ$C. 

To rule out that this surprising observation does not originate from the different layer thickness of 20\,nm in our case compared with a thickness of 3\,nm in most magnetic tunnel junctions \cite{Ikeda:2008p3496}, consequently, we prepared another sample with an Al$_{95}$Si$_5$ (4.0\, nm)/MgO (2.1\,nm)/ Co$_{40}$Fe$_{40}$B$_{20}$ (3\,nm)/ Ta (10\,nm) layer stack. The additional tantalum layer improves the electrode film conductivity which otherwise leads to so-called ``current crowding'' effects modifying the transport measurements \cite{Moodera:1995p66}, which could influence the peak heights \cite{Worledge:2000p55}. 

\begin{figure}\hfill
	\includegraphics{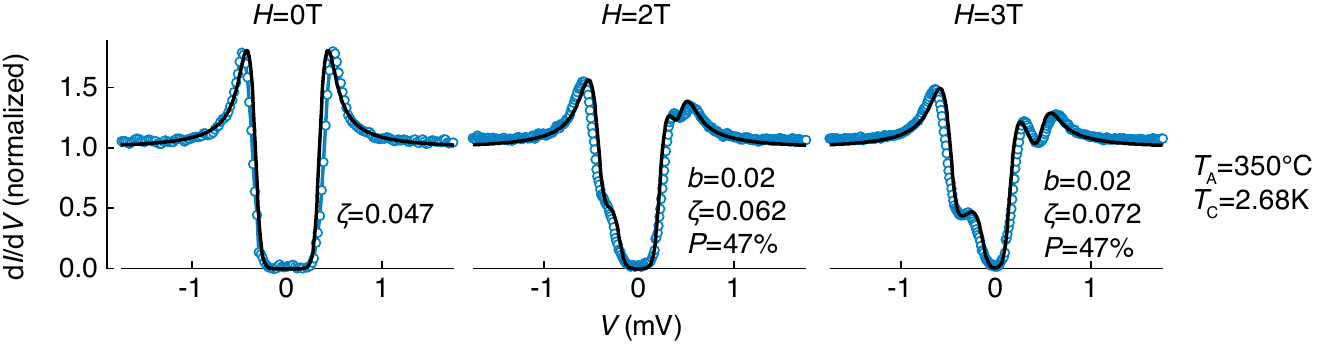}
	\caption{Differential conductance of a Al$_{95}$Si$_5$ (4.0\,nm)/ MgO (2.1\,nm)/ Co$_{40}$Fe$_{40}$B$_{20}$ (3\, nm)/ Ta (10\,nm) junction in different magnetic fields. The experimental data are given in blue, and the Maki fit is given in black.}
	\label{CoFeB-ta_SPT}
\end{figure}

\begin{figure}\hfill
	\includegraphics{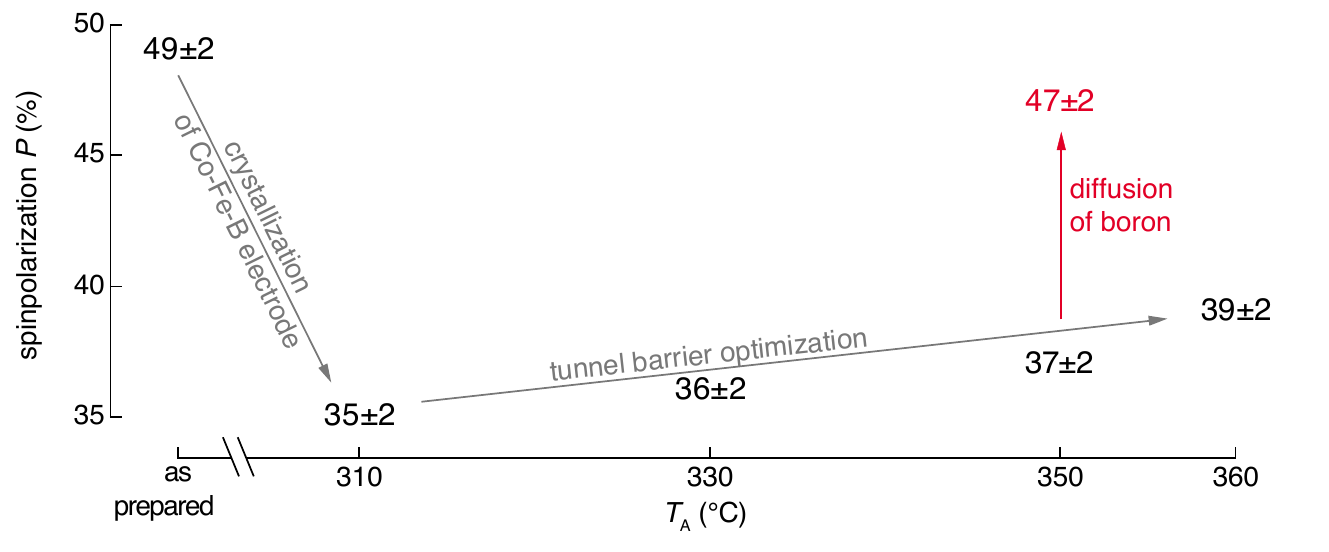}
	\caption{The spin polarization values as a function of the annealing temperature. The arrows indicate the effect of the three different mechanisms explained in the text and their origin.}
	\label{CoFeB_SpinPol}
\end{figure}

The d$I/$d$V$ measurements of these junctions that were annealed at $T_a=350\,^\circ$C are depicted in Figure~\ref{CoFeB-ta_SPT}. In this sample, the spin polarization was increased and showed values of 47$\pm$2\%, although all other parameters, such as $\zeta$ and $b$, were approximately the same. To explain the results, we divided the spin polarization curve depicted in Figure \ref{CoFeB_SpinPol} into three parts: the initial drop at $T_a =  310\,^\circ$C, the slow increase with further increases in annealing temperature and the higher values of the Co-Fe-B/ Ta double layers.  

We began with the initial drop of the spin polarization and compared our results to those from alumina-based junctions. We disregarded symmetry filtering effects, which are observed at higher temperatures of more than $T_a =  350\,^\circ$C \cite{Parkin:2004p71}. Gao et al.\ have investigated the spin polarization in Al$_{95}$Si$_5$ /Al$_2$O$_3$ /Co-Fe \textsc{mtj}s \cite{Gao:2009p8110}. They found a spin polarization of amorphous, ultra-thin Co-Fe electrodes of $P = 55.2\%$, which exceeds the polarization values, $P = 39.1 \%$, of the annealed samples. This behavior has also been observed in magnetic tunnel junctions \cite{Gao:2009p10988} and corresponds well with the initial drop of $P$ observed in our experiments. 

To investigate the direct influence of the boron aside from the amorphization of the Co-Fe, Paluskar et al.\ determined the spin polarization of Co$_{72}$Fe$_{20}$B$_8$ electrodes at different layer thickness and annealing temperatures  \cite{Paluskar:2008p23}. They compared the experimental results with calculations based on density functional theory calculations of electrodes with either a bcc crystal structure or amorphous constitution. In their calculations and in their experiments, Paluskar et al.\ also found a lower spin polarization of the tunneling electrons in the junctions with crystalline electrodes compared with junctions with amorphous layer electrodes. If we apply their calculations to our composition of Co$_{40}$Fe$_{40}$B$_{20}$, we obtain an expected polarization value of $P=50.6$\%. This matches our experiments which yield a polarization value of $P = 49\pm2$\%

Next, we examined the slow but continuous increase of the spin polarization with increasing annealing temperatures from $T_a=310\,^\circ$C to $T_a=360\,^\circ$C, which is also depicted in Figure \ref{CoFeB_SpinPol}. Similar experiments with alumina-based junctions have yielded no increase, even for annealing temperatures of up to $T_a=500\,^\circ$C \cite{Kant:2004p10982}. Conversely, MgO-based tunnel junctions show a crystallization of the electrode/ MgO interface and, therefore, an increase of the spin polarization of the tunneling electrons due to symmetry filtering \cite{Parkin:2004p71}. This increase has also been observed in \textsc{mtj}s with Co-Fe-B electrodes and MgO tunnel barriers: a \textsc{tmr}-value of up to 600\% has been determined at room temperature \cite{Ikeda:2008p3496}. In the present study, the increase in this work can also be attributed to a crystallization of the electrode-barrier interface, although our sample preparation method does not allow for the high annealing temperatures that can lead to polarization values beyond 80\% \cite{Parkin:2004p71}.  

The mechanism responible for the higher spin polarization in the junctions with Co-Fe-B/Ta electrodes, as shown in Figure \ref{CoFeB_SpinPol}, is still an open question. Investigations of Co-Fe-B/MgO/Co-Fe-B-\textsc{mtj}s with hard x-ray photo emission spectroscopy (\textsc{haxpes}) have shown that boron diffusion in the adjacent tantalum layer is responsible for the high \textsc{tmr}-ratios \cite{Kozina:2010p10990}. Although higher annealing temperatures of $T_a =  500\,^\circ$C were used in the \textsc{haxpes} experiments,  the results suggest that the same mechanism might be responsible for the increase of the tunneling spin polarization in our work. 

\section{Co$_2$FeAl-based junctions}
Apart from increasing the tunneling spin polarization by symmetry filtering, a second approach to targeting higher \textsc{tmr} ratios is by maximization of the  spin polarization of the ferromagnetic electrodes. Half-metallic ferromagnets exhibit the highest possible polarization of 100\% \cite{DeGroot:1983p3287}; this is predicted for CrO$_2$ \cite{Schwarz:1986p10273}, magnetite \cite{Penicaud:1992p13934} and for Heusler compounds. While the former compounds are fixed in their composition and constituents, the Heusler compounds can be tailored by substitution of the constituents \cite{Balke:2008p8665, Ebke:2006p14327}. 

As shown by Ebke et al., Co$_2$FeAl has a low crystallization temperature and yields high \textsc{tmr} ratios when included as a ferromagnetic electrode in \textsc{mtj}s \cite{Ebke:2010p6784}. Therefore, the Heusler compound Co$_2$FeAl has been investigated. Ebke et al have reported that the Heusler compounds in \textsc{mtj}s require a particular layer stack. The electrode had to be grown on a 5\,nm MgO buffer on MgO substrates \cite{Ebke:2010p6784}. In our case, this requires an inverted layer stack with the ferromagnetic electrode at the bottom and the superconductor at the top of the system. This inversion and the MgO buffer layer and substrate demand a different thickness of the Al-Si to ensure good superconducting properties of the spin detector. This has also been reported by Yang et al.\ \cite{Yang:2007p35}. We used a layer stack of MgO (5\,nm)/ Co$_2$FeAl (20\,nm)/ MgO (2.1\,nm)/ Al$_{95}$-Si$_5$ (5\,nm) and investigated annealing temperatures of $350\,^\circ$C and $375\,^\circ$C, where the largest change and highest \textsc{tmr} ratios were found in our earlier experiments \cite{Ebke:2010p6784}.

\begin{figure}\hfill
	\includegraphics{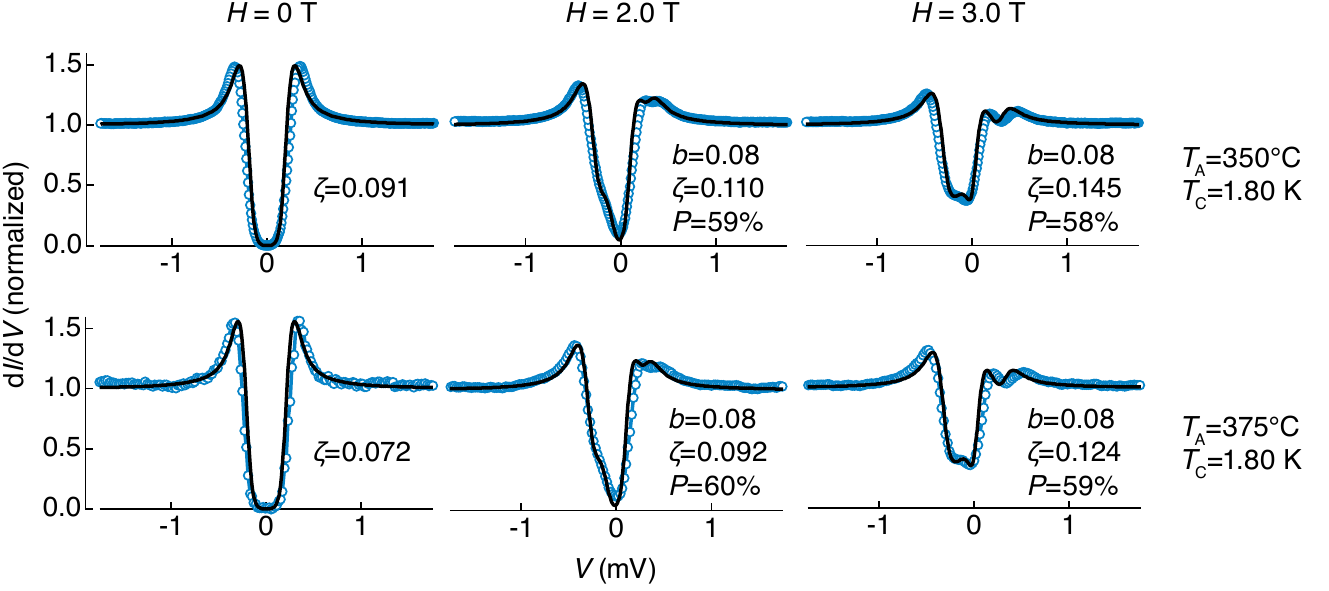}
	\caption{Differential conductance of MgO (5nm)/Co$_2$FeAl (20nm)/MgO (2.1nm)/AlSi (5nm) tunnel junctions. The measurements reveal a spinpolarization of $59$\,\% for both annealing temperatures.}
	\label{CoFeAl_SPT}
\end{figure}

In Figure~\ref{CoFeAl_SPT}, the experimental and the theoretical $\mathrm{d}I/\mathrm{d}V-V$ curves of the Co$_2$FeAl-junctions are shown. From the fit of the theoretical to the measured curves, a spin polarization of $59\pm2$\% can be deduced for both the sample annealed at $350\,^{\circ}$C and the sample annealed at $375\,^{\circ}$C. This is similar to the value $P=56.2$\,\% given by Innomata, which was calculated from \textsc{tmr} values of Co$_2$FeAl/Al$_2$O$_3$/Co-Fe \textsc{mtjs} measured at $T=5$\,K \cite{Inomata:2008p3157}. In the same study, a theoretical value of $P = 60.7$\,\% for this compound was reported, which is also in good agreement with the value obtained in our experiments. Compared to the Co- and Co-Fe-B tunnel junctions, higher spin-orbit scattering and pair-breaking values had to be used, as previously seen by by Yang et al. \cite{Yang:2007p35} for superconducting Al-Si-electrodes on top of an MgO tunnel barrier.

Because of the good agreement between the measured and theoretically predicted spin polarizations, the increase in the \textsc{tmr} ratio can be attributed to symmetry filtering of the tunnel barrier. Additionally, the oscillation of the \textsc{tmr} ratio with increasing MgO thickness in Co$_2$FeAl/MgO/Co-Fe-\textsc{mtj}s that has been found by Wang \cite{Wang:2010p8710} et al. suggests that the high \textsc{tmr} ratios in \textsc{mtj}s containing Co$_2$FeAl are based upon symmetry filtering rather than highly spin-polarized electrodes. 
\section{Summary}
We investigated the transport properties of tunnel junctions with a magnetic electrode and a superconducting Al-Si counter electrode with and without magnetic fields for different post annealing temperatures. This included the measurement of spin polarization by Meservey-Tedrow technique. We evaluated the spin polarization as well as the spin orbit scattering and total pair breaking parameter using the Maki-Fulde-model, and we compared the data with the corresponding magnetic tunnel junctions. These two techniques complement each other to provide a better understanding of the underlying physics. In the future, samples with superconducting counter electrodes and prepared with a lithographic process similar to that used to produce magnetic tunnel junctions would be useful.
\ack{O.\ Schebaum and A.\ Thomas acknowledge the MIWF of the NRW state government for financial support. J.S.\  Moodera acknowledges the financial support from NSF grant DMR 0504158 and ONR grant N00014-09-1-0177.}
\section*{References}
\bibliography{bib}
\bibliographystyle{unsrt}

\end{document}